# Symmetry-Preserving Local Representation for Short-Range Interaction in Complex Materials

Hao Cheng,[1 †] Tong Yang,[1, †] Hao Wu,[1] Haokai Hong,[2] Zhongrui Wang,[3] Shu Ping Lau,[1] Wanyu Lin,[2, *] Kay Chen Tan,[4, *] Ming Yang[1, 5, 6, *]

[1] Department of Physics and Materials, The Hong Kong Polytechnic University, Kowloon, Hong Kong SAR, China

[2] Department of Computing, The Hong Kong Polytechnic University, Kowloon, Hong Kong SAR, China

3 The School of Microelectronics, Southern University of Science and Technology, Shenzhen, China

[4] Department of Data Science and Artificial Intelligence, The Hong Kong Polytechnic University, Kowloon, Hong Kong SAR, China

[5] Research Centre on Data Sciences & Artificial Intelligence, The Hong Kong Polytechnic University, Kowloon, Hong Kong SAR, China

[6] Research Centre for Nanoscience and Nanotechnology, The Hong Kong Polytechnic University, Kowloon, Hong Kong SAR, China

[†] These authors contributed equally to this work.

[*]Correspondence and requests for materials should be addressed to Ming Yang (kevin.m.yang@polyu.edu.hk), Wanyu Lin (wan-yu.lin@polyu.edu.hk), or Kan Chen Tan (kaychen.tan@polyu.edu.hk).

**ABSTRACT** Modeling short-range interactions in highly disordered materials remains challenging due to the immense configuration space and the disruption of local symmetry in truncated theoretical models. Here, we propose a symmetry-preserving local representation framework that strictly enforces topological continuity and translational invariance via boundary passivation. Applied to high-entropy alloys, this approach autonomously unravels the nonlinear ligand effects modulating local reactivity and accurately captures coverage-dependent many-body interactions without explicit periodic expansion. By resolving the conflict between localized chemical environments and extended periodic boundaries, this framework not only reduces computational complexity by an order of magnitude but also provides a physically transparent paradigm for deciphering short-range order in complex condensed matter systems.

## I. INTRODUCTION

The physical properties of many complex condensed matter systems, ranging from the catalytic activity of high-entropy alloys (HEAs) [1-4] and the defect physics of low-dimensional materials [5-7], to the dynamic solvation processes at solid-liquid interfaces [8-10], are dominantly governed by their local chemical environments and short-range order (SRO). Across these highly diverse systems, the immense combinatorial space of local atomic arrangements and structural symmetry-breaking lead to a broad distribution of site-specific behaviors [11,12]. These behaviors are driven by intricate, short-range many-body interactions, such as nonlinear ligand effects in alloys or explicit solvent perturbations at interfaces, where the local coordination shell dynamically modulates the electronic structure of the active sites [13-15]. Understanding and systematically predicting these local structure-property mappings under complex local coordination remains challenging in modern solid-state physics and materials science [16,17].

Recent advances in machine-learning potentials and graph neural networks (GNNs) have revolutionized atomistic modeling by bypassing explicit electronic structure calculations [18-21]. However, modeling complex localized phenomena presents a fundamental trade-off between physical fidelity and computational scalability. On one hand, high-fidelity approaches, such as full-crystal message-passing networks [22-27], inherently preserve the topological continuity of the lattice and seamlessly capture extended many-body interactions without artificial truncation. While these models successfully overcome the steep $O(n^3)$ scaling of density functional theory (DFT) calculations, they still suffer from computational complexity, ranging from an $O(n)$ scaling for deep local message-passing to an intractable $O(n^2)$ bottleneck for global architectures [28-31]. Consequently, its computational cost remains high to rapidly explore the vast structural landscapes of disordered alloys or large-scale dynamic electrochemical interfaces using full-crystal graph representation [10,32]. On the other hand, local structure-based representations, such as atom-centered symmetry functions and local subgraphs, have been proposed to drastically reduce the computational burden [33-38]. Nevertheless, simple spatial truncation in these local models artificially disrupts interatomic bonds at the cluster boundary. This rigid truncation breaks the translational symmetry of the crystal and destroys the local topological coordination. Inevitably, it induces boundary artifacts that propagate non-physically during iterative information aggregation, fundamentally obscuring the true short-range chemical interactions [39,40].

In this Letter, we resolve this conflict by developing Cluster Graph Neural Network (CG-NET), a symmetry-preserving local representation framework that rigorously maintains topological continuity and periodic consistency without requiring lattice expansion. By introducing a boundary passivation scheme utilizing topological pseudo-nodes alongside a symmetry-weighted sampling strategy, we mathematically compensate for truncated coordination environments and restore translational invariance within a highly localized graph. Applying this framework across various complex systems, including highly disordered HEAs, 2D impurity matrices, metal-organic frameworks, and explicit solid-liquid interfaces, we demonstrate its capability to extract underlying physical laws. Remarkably, direct analysis of the learned latent space autonomously unveils the spatial dependency of nonlinear ligand effects. Furthermore, the symmetry-preserving local environment naturally captures coverage-dependent adsorbate-adsorbate interactions, offering a physically transparent and efficient paradigm for deciphering localized physical phenomena in extended solid-state materials.

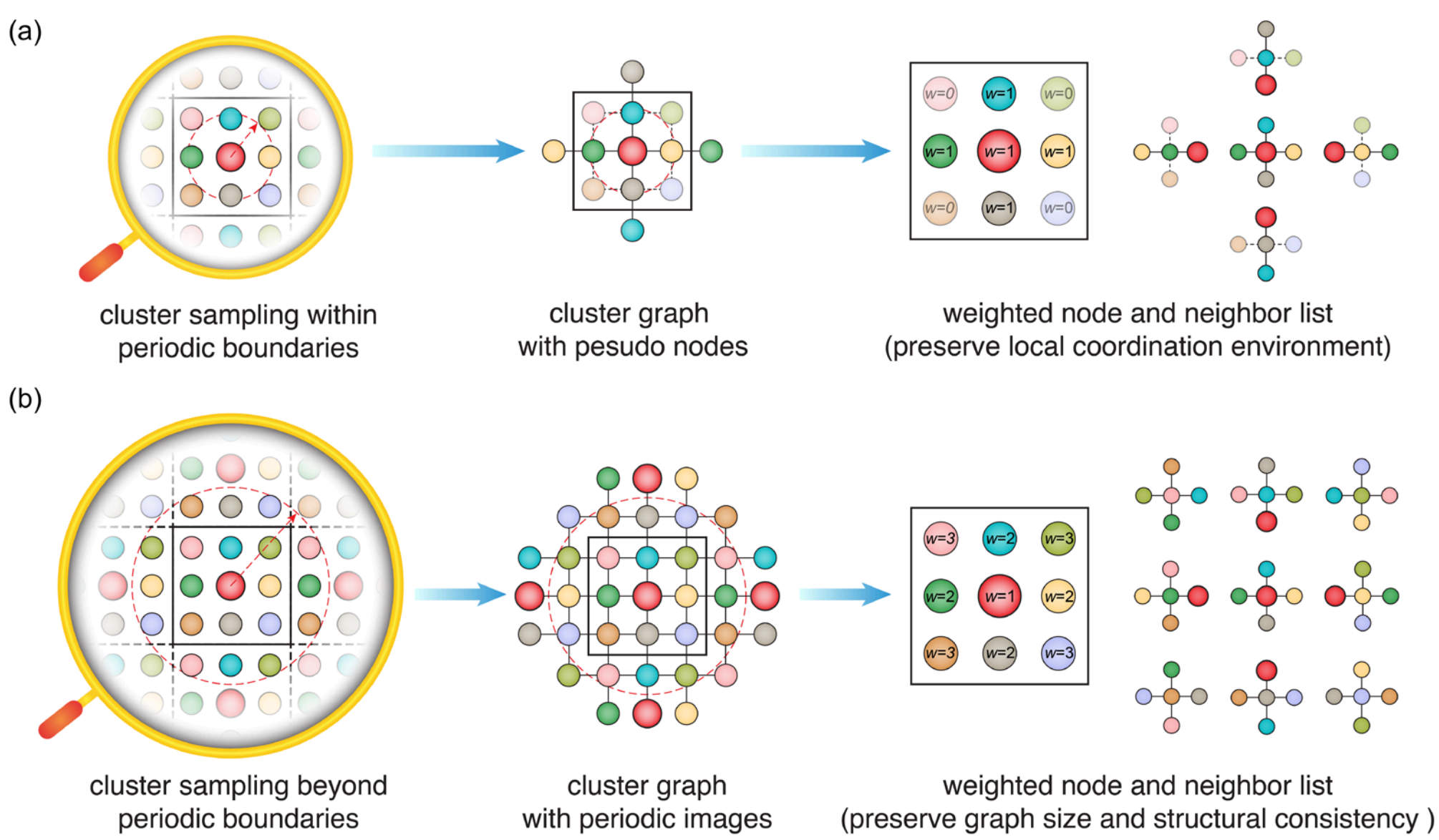

FIG. 1. Symmetry-preserving localized graph representation. (a) Topological boundary passivation using pseudo-nodes (shaded) to restore the coordination environment of truncated boundary atoms. (b) Enforcing translational invariance by mapping periodic images back to the primitive cell using multiplicity weights, folding extended boundary conditions into a compact local basis.

## II. METHODS

To model these localized phenomena, we conceptualize the local structural environment as an atomic graph centered on a target motif (e.g., a catalytic active site or a point defect) confined within a physical cutoff radius $R_c$. In conventional localized graph representations, atoms located near the truncated boundary ($r \approx R_c$) suffer from artificial coordination loss. Physically, this spatial truncation is akin to creating dangling bonds at a surface, which drastically perturbs the effective local density of states and introduces spurious boundary scattering during iterative information aggregation [41]. Such boundary-induced artifacts distort the representation of the local chemical environment, severely degrading both the prediction accuracy and physical reliability of the model.

To overcome this limitation, we introduce a topological boundary passivation scheme alongside a symmetry-weighted sampling strategy, as illustrated in Fig. 1. For boundary atoms whose complete coordination shells are truncated by $R_c$, pseudo-nodes are added to complete their crystalline coordination [Fig. 1(a)]. Crucially, these pseudo-nodes act strictly as topological boundary conditions rather than independent physical variables. They carry no learnable degrees of freedom, yet they seamlessly provide the missing geometric placeholders required for correct spatial convolutions. By artificially saturating the broken bonds at the truncated boundary, this passivation scheme preserves the exact local coordination topology, ensuring that the chemical environment of boundary atoms remains physically consistent with the extended periodic boundary.

Furthermore, capturing long-range spatial correlations (such as those governing van der Waals interaction or high-coverage adsorbate interactions) often necessitate a cutoff radius that extends beyond the primitive unit cell. Explicitly replicating periodic image atoms in the graph drastically inflates the computational costs and breaks the uniqueness of the primitive basis. Instead, we enforce a translational invariance by mapping all symmetry-equivalent periodic images back to their corresponding parent atoms within the primitive cell [Fig. 1(b)]. We encode this periodicity by assigning a mathematical multiplicity weight to each node, effectively folding the extended periodic boundary conditions into the structural weights of the localized graph.

Mathematically, let $\mathcal{V} = \{v_1, v_2, \dots, v_N\}$ denote the global set of atoms in the extended crystal, and $\mathcal{V}_{prim} \subset \mathcal{V}$ denote the unique set of parent atoms within the primitive unit cell. To enforce translational invariance, the active atoms comprising the localized graph $\mathcal{V}_{local}$ are strictly restricted to this primitive basis. Thus, $\mathcal{V}_{local}$ consists of primitive atoms that have at least one periodic image falling within a cluster radius $R_c$ centered at an origin $\boldsymbol{r}_0$. This is defined by a multiplicity weight $w_i = \sum_{\boldsymbol{R}_n} 1\,(\|\boldsymbol{R}_i + \boldsymbol{R}_n - \boldsymbol{r}_0\| \le R_c)$ where $\boldsymbol{R}_n$ represent a lattice translation vector and $1(\cdot)$ is the indicator function. Consequently, the localized graph is defined as $\mathcal{V}_{local} = \{v_i \in \mathcal{V}_{prim} \middle| w_i \ge 1\}$. For any atom $v_i \in \mathcal{V}_{local}$, its complete local coordination neighborhood within a message-passing cutoff $r_c$ is $\mathcal{N}(i) = \{j \in \mathcal{V} \middle| \|\boldsymbol{R}_j - \boldsymbol{R}_i\| \le r_c,\ j \ne i\}$. This full neighborhood is explicitly partitioned into two disjoint sets. Conventional localized models exclusively consider the internal neighborhood, $\mathcal{N}_{local}(i) = \{j \in \mathcal{N}(i) \middle| \|\boldsymbol{R}_j - \boldsymbol{r}_0\| \le R_c\}$, which contains only the neighbors fully retained within the localized graph boundary. By ignoring atoms outside this boundary, traditional methods inevitably suffer from artificial coordination loss. In contrast, our approach incorporates the missing boundary environment through the passivation shell, $\mathcal{N}_{pseudo}(i) = \{p \in \mathcal{N}(i) \middle| \|\boldsymbol{R}_p - \boldsymbol{r}_0\| > R_c\}$, which comprises the truncated neighbors falling outside the boundary. The symmetry-preserving spatial convolution for updating the hidden state $h_i^t$ of a target atom $i$ can then be formalized as:

$$m_i^t = \sum_{j \in \mathcal{N}_{local}(i)} M_t(h_i^t, h_j^t, \boldsymbol{r}_{ij}) + \sum_{p \in \mathcal{N}_{pseudo}(i)} M_t(h_i^t, h_p^{fixed}, \boldsymbol{r}_{ip}) \quad (1)$$

$$h_i^{t+1} = U_t(h_i^t, m_i^t) \quad (2)$$

where $M_t$ is the message function, $U_t$ is the state update function, and $\boldsymbol{r}$ denotes the relative distance vector. Crucially, $h_p^{fixed} = \mathbf{e}(Z_p)$ where $\mathbf{e}(Z_p)$ is a deterministic, non-learnable embedding corresponding to the atomic number $Z_p$ of the truncated boundary atom. Finally, the physical observable $y$ is predicted during the readout phase by pooling the final hidden states $h_i^T$ over the localized graph using a readout function:

$$\hat{y} = R(\{w_i h_i^T | v_i \in \mathcal{V}_{local}\}) \quad (3)$$

Here, scaling the atomic $w_i$ contributions by the multiplicity weight effectively folds the extended periodic boundary conditions into the localized graph, ensuring that the global prediction preserves exact translational invariance.

Together, these boundary passivation and symmetry-weighted sampling mechanisms provide a powerful structural inductive bias. Unlike conventional data-driven models that implicitly learn the rules of periodicity and bonding through massive datasets, our formulation inherently respects the spatial symmetries, local topological continuity, and translational invariance of the crystal lattice. To validate our CG-NET framework, we employ the Universal Models for Atoms (UMA) [27] as the equivariant message-passing backbone. Importantly, this symmetry-preserving formalism is agnostic to the underlying neural network, providing a local basis for exploring solid-state interactions at a fraction of the traditional computational cost.

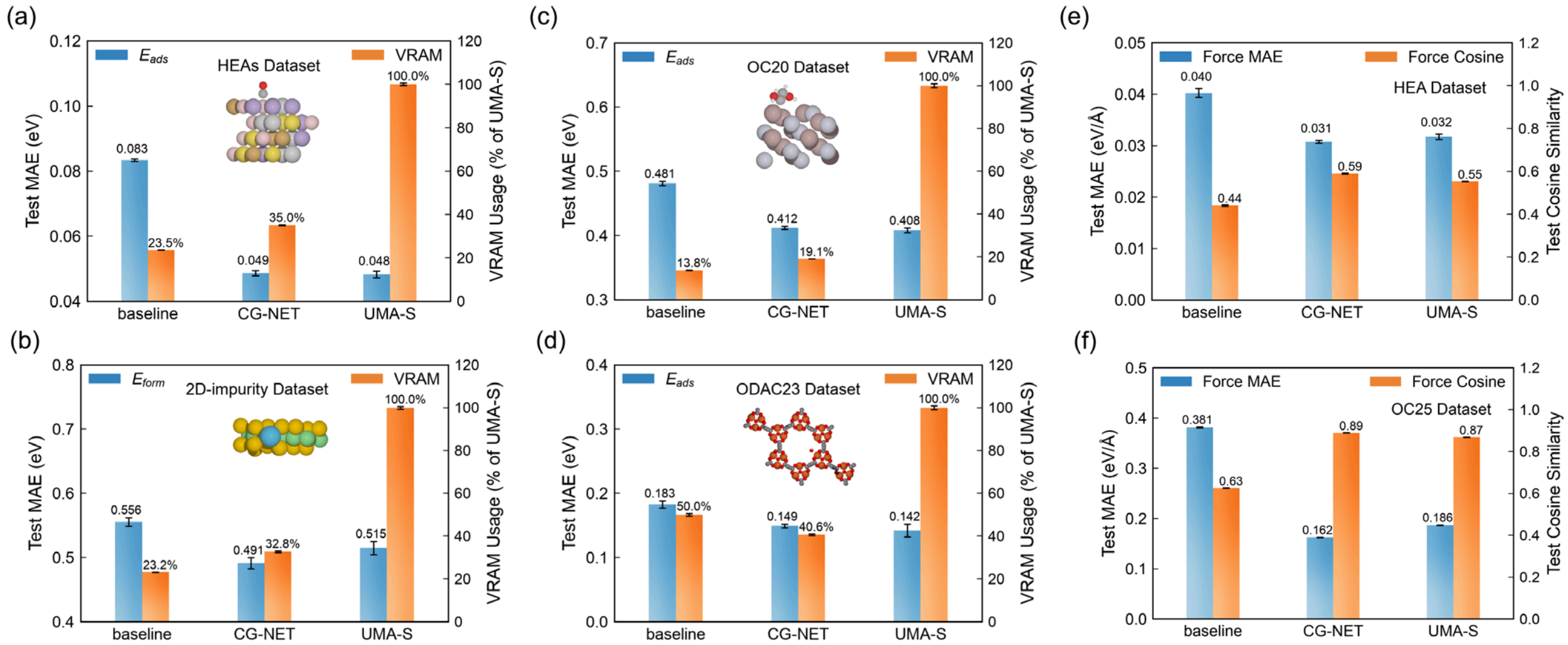

FIG. 2. Prediction fidelity of the symmetry-preserving local representation across diverse physical regimes. (a-d) Mean absolute errors (MAEs) and GPU video memory (VRAM) consumption for adsorption/formation energy predictions spanning highly disordered alloys (HEA), point defects (2D-impurity), diverse catalytic surfaces (OC20), and weak van der Waals frameworks (ODAC23). (e-f) MAEs and cosine similarities for conservative atomic forces, probing structural disorder (HEA) and complex interfacial solvation dynamics (OC25). To ensure a fair comparison, identical local graph sizes were maintained for both the baseline and CG-NET models across all evaluations. In all panels, performance is compared across an unpassivated localized model (baseline), our proposed topologically passivated framework (CG-NET), and a state-of-the-art global graph network (UMA-S).

## III. RESULTS

Before interrogating the underlying physical mechanisms, we first establish the prediction fidelity and universality of this symmetry-preserving framework. The model was optimized to predict adsorption energy, formation energy, and conservative atomic forces across fundamentally distinct physical regimes. These encompass strong covalent and metallic bonding in highly disordered HEA surfaces [see Sec. 1 in Supplementary Material (SM)] and 2D impurity [42]; vast compositional spaces of generic catalytic surfaces (OC20) [43]; weak van der Waals interactions in complex metal-organic frameworks (ODAC23) [44]; and expansive solvent dynamics at explicit solid-liquid interfaces (OC25) [10] [see Sec. 2 in SM].

Remarkably, as summarized in Fig. 2, simple spatial truncation (the unpassivated baseline) severely degrades predictive accuracy due to disrupted coordination topologies. For instance, on the HEA dataset, the baseline yields an energy mean absolute error (MAE) of 0.083 eV and a force MAE of 0.040 eV/Å [Fig. 2(a) and (e)]. By integrating topological pseudo-nodes, CG-NET significantly rectifies these distortions, reducing the errors to 0.049 eV and 0.031 eV/Å, respectively. This performance matches that of the computationally heavy UMA-S model, which yields 0.048 eV and 0.032 eV/Å. Consistently across all other datasets, our localized representation continues to rival the state-of-the-art global graph network, particularly excelling in complex interfacial force predictions (0.162 eV/Å vs. 0.186 eV/Å on OC25) [Fig. 2(f)].

Beyond prediction accuracy, our CG-NET resolves the computational bottlenecks associated with extended crystalline modeling. Explicitly replicating periodic images in full-lattice graphs cubically inflates graph dimensions and computational overhead. By mapping these periodic redundancies into mathematical node weights, our approach reduces the required graphical edges by over 77% to 92% across the different datasets [see Table S1 in SM]. As illustrated in Fig. 2(a-d), this structural compactness translates to a remarkable and systematic reduction in GPU video memory (VRAM) consumption across all datasets, ranging from approximately 60% to over 81% relative to the global graph model. This profound efficiency advantage confirms that local symmetry preservation is mathematically sufficient to reconstruct physical observables without the immense computational overhead of full-lattice aggregation.

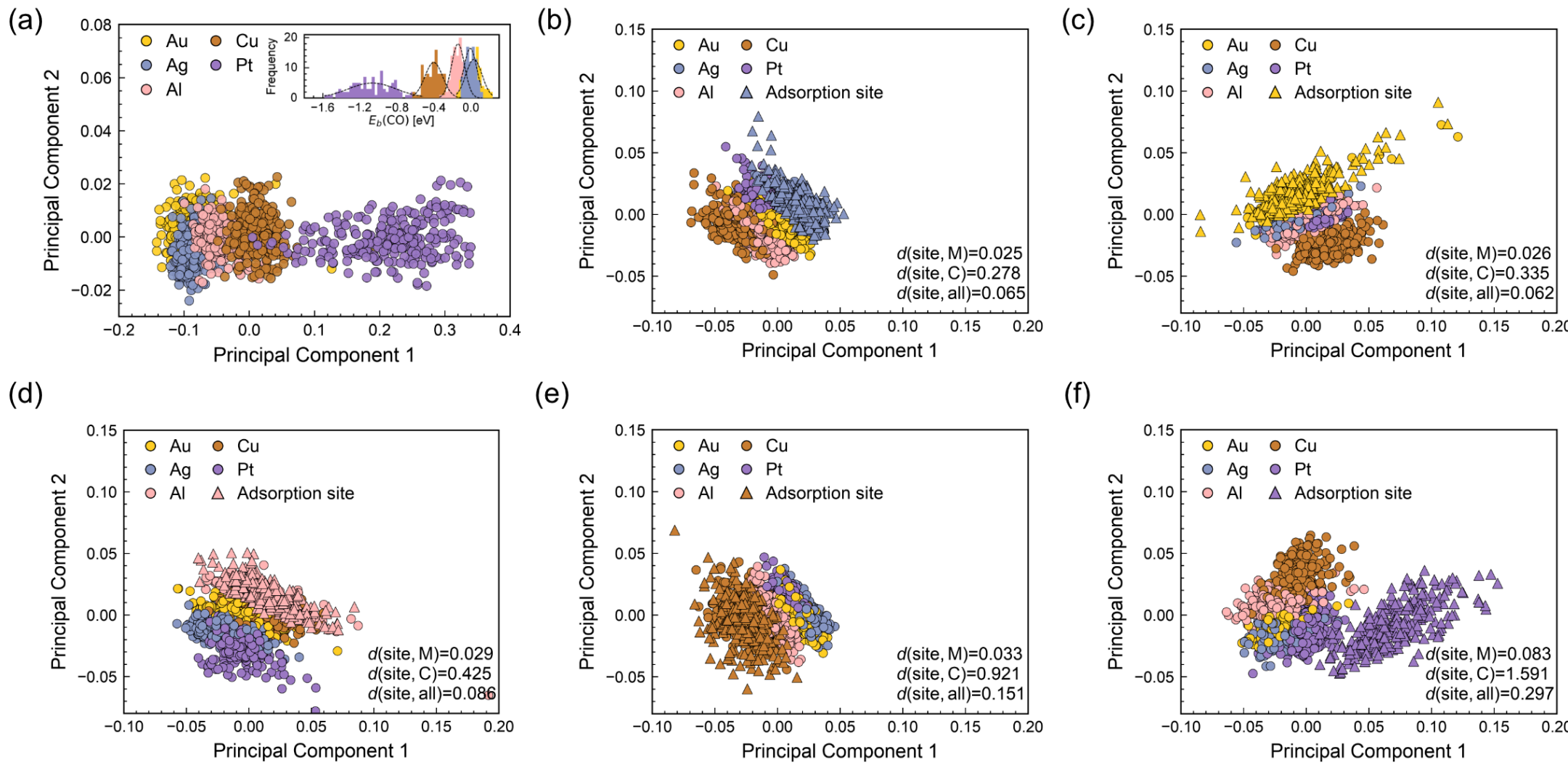

FIG. 3. Data-driven extraction of short-range ligand effects and reactivity trends in the AgAuAlCuPt high-entropy alloy. (a) Principal component analysis (PCA) of the topologically passivated latent space, revealing the spontaneous segregation of element-specific thermodynamic domains for *CO adsorption. The inset validates this representation against ab initio adsorption energy distributions. (b–f) Latent-space mapping of the localized coordination environments for the five distinct metallic active sites (triangles) and their surrounding neighborhood (circles). The spatial metric in this latent representation, $d_{(\text{site, all})}$, inherently quantifies the cumulative electronic perturbation from the local environment, strictly recovering the fundamental reactivity hierarchy (Pt > Cu > Al > Ag ≈ Au) without explicit physical priors.

With the established prediction capabilities, we further seek to uncover the physical origins of this high fidelity by probing the complex surface of the AgAuAlCuPt HEA. Rather than relying on empirical physical descriptors, we directly investigate the high-dimensional latent space formulated by the symmetry-preserving framework. As illustrated in Fig. 3(a), the principal component analysis (PCA) of the latent atomic embeddings spontaneously segregates into element-specific thermodynamic domains, perfectly mirroring the theoretically calculated adsorption energy distributions [see. Fig. S1 in SM]. Notably, the Pt embeddings exhibit a remarkably broader and more distinct distribution compared to the localized domains of other constituent metals (Ag, Au, Al, and Cu). This data-driven segregation directly reflects the heightened sensitivity of Pt to local structural perturbations, arising from the relatively higher energy position of its unfilled d-orbitals near the Fermi level [45,46]. This confirms that the structurally constrained local graph inherently resolves the fundamental electronic distinctions of heterogeneous binding sites.

More profoundly, the representation autonomously disentangles the intricate short-range many-body interactions between the active site and its coordination shell [Fig. 3(b-f)]. By evaluating the inter-cluster distance $d_{(\text{site, all})}$ between the binding site and its neighbors in the latent space, a topological metric quantifying the cumulative electronic perturbation exerted by the local environment, we observe a strict hierarchical ordering of element-specific reactivity in the HEA: Pt (0.297) > Cu (0.151) > Al (0.086) > Ag (0.065) ≈ Au (0.062). Strikingly, this latent geometric metric exhibits a near-perfect monotonic correlation (Spearman rank $\rho$ = -0.90) with the adsorption energies [see Fig. S2 in SM]. However, this physical trend is severely obscured in both unpassivated local models (baseline, $\rho$ = 0.22), which fundamentally distort the coordination topology [see Fig. S3 in SM], and global graph networks (UMA-S, $\rho$ = -0.37), where localized ligand effects are diluted by unconstrained long-range convolutions [see Fig. S4 in SM]. Ultimately, our topologically passivated local representation can discover the spatial dependency of the nonlinear ligand effect, providing a direct, data-driven manifestation of how short-range structural perturbations dictate chemical reactivity in the presence of extreme compositional disorder.

Beyond the localized reactivity of isolated sites, a rigorous theoretical framework must also correctly capture extended many-body phenomena, such as coverage-dependent lateral interactions. These adsorbate-adsorbate repulsions dictate thermodynamic phase transitions and overall catalytic reaction rates [47,48]. We evaluate this capability by training exclusively on an intermediate adsorbate coverage (3×3 supercells) and forcing the model to extrapolate to dense (2×2) and sparse (4×4) interacting regimes [Fig. 4]. In conventional global graph networks, altering the supercell dimensions fundamentally changes the graph size and message-passing pathways, leading to representation collapse. In contrast, because our topological passivation scheme strictly anchors the representation to the invariant primitive cell, it inherently encodes these spatial many-body interactions without requiring expanded basis sets. Consequently, our local framework maintains exceptional physical consistency across untrained boundary conditions, yielding highly robust $R^2$ correlation coefficients (0.94 and 0.97) and low MAEs (0.073 eV and 0.058 eV). This demonstrates that extended lateral interactions can be accurately captured within a strictly localized representation, providing that the underlying translational symmetry is rigorously maintained.

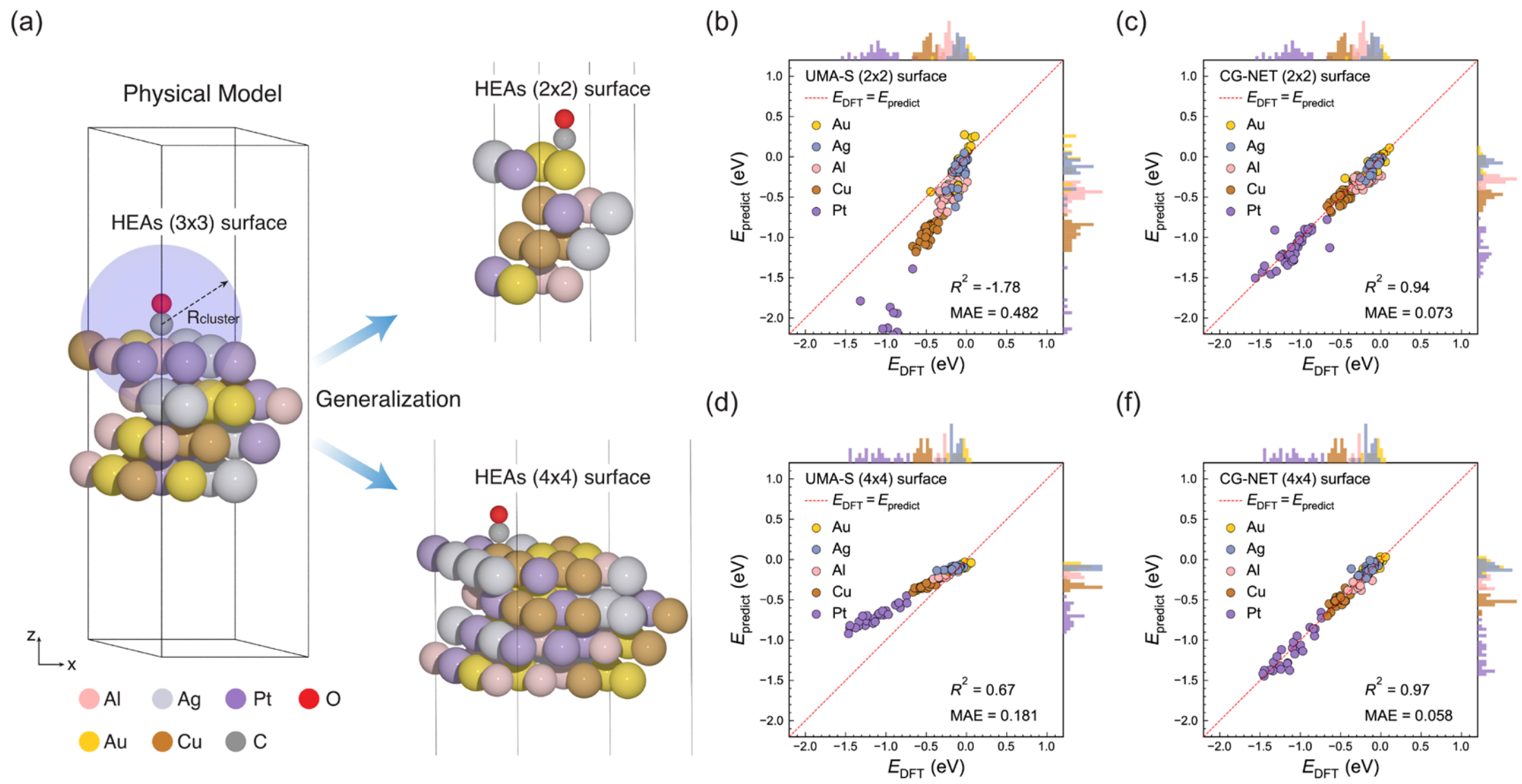

FIG. 4. Capturing coverage-dependent many-body interactions across varying coverage regimes. (a) Schematic illustrating the phase-space extrapolation: models trained exclusively on an intermediate adsorbate coverage (3×3 HEA supercells) are evaluated on highly dense (2×2) and sparse (4×4) interacting regimes. (b–e) Predictive performance of the symmetry-preserving local representation (CG-NET) compared to a conventional full-crystal counterpart UMA-S model under varying lateral coverages.

## IV. CONCLUSIONS

In summary, we have bridged the gap between extended periodic symmetry and localized chemical environments by introducing a topologically passivated, symmetry-preserving representation. By utilizing deterministic pseudo-nodes and periodic multiplicity mapping, this framework circumvents the boundary truncation artifacts that have historically hindered localized models. It restores exact spatial symmetries without the prohibitive dimensional explosion inherent to full-lattice calculations. Beyond achieving universal prediction fidelity across diverse bonding regimes, this structurally constrained representation natively unveils the underlying physics of disordered interfaces. It directly extracts the spatial dependence of nonlinear ligand effects from its latent geometry and inherently captures many-body lateral interactions across varying coverage regimes. Ultimately, by embedding physical inductive biases into the localized graph topology, our approach provides a highly interpretable and computationally efficient paradigm. By effectively mitigating traditional scaling limitations, this framework offers a highly scalable approach for exploring localized quantum chemical phenomena, extreme structural disorder, and complex dynamic interfaces in condensed matter physics.

## DATA and CODE AVAILABILITY

The training data for the CG-NET are available at https://doi.org/10.6084/m9.figshare.29413064. The source code for CG-NET is available at https://github.com/hchenglab/CG-NET.

## ACKNOWLEDGMENTS

This work was supported by Hong Kong Polytechnic University (project number: P0042711, P0051269, P0050570, P0058378, P0058408, P0058414, P0064643, and P0048122), and Hong Kong Research Grants Council (project number: 15307124).

## AUTHOR CONTRIBUTIONS

M. Y. and H. C. conceived the idea. M. Y., W. L., and K. C. T. designed the study. H. C., T. Y. implemented the code. H. C., H. W., H. K. H., Z. R. W., S. P. L., W. L., K.C.T., and M. Y. prepared the dataset and performed the analysis. H.C. drafted the manuscript. All the authors have read and corrected the final manuscript.

## COMPETING INTERESTS

The authors declare no competing interests.